\documentclass[twocolumn]{article} 
\usepackage{times}
\usepackage{graphicx}
\usepackage{amssymb}
\usepackage{url,hyperref}

\usepackage{setspace}
\usepackage{algorithm}
\usepackage{algorithmic}
\usepackage{amssymb}
\usepackage[figuresright]{rotating}
\usepackage{graphicx}
\usepackage{amssymb}
\usepackage{epstopdf}		
\usepackage{caption}
\usepackage{xcolor}

\usepackage{subcaption}

\usepackage{amsmath,amsfonts,amsthm} % Math packages

\usepackage{graphicx}
\usepackage{epstopdf}

\usepackage{float}

\newcommand{\titolo}{ A study on users' privacy perception with smart devices.}

\begin{document}
\title{\titolo}

\author{Alan Ferrari, and Silvia Giordano \\
\\Institute for Information System and Networking\\University of Applied Sciences of Southern Switzerland (SUPSI)\\
Manno, Switzerland\\
firstname.lastname@supsi.ch
}
%% author names and affiliations
%% use a multiple column layout for up to three different
%% affiliations
%\author
%{
%%\IEEEauthorblockN{Alan Ferrari, Daniele Puccinelli, and Silvia Giordano}
%%\IEEEauthorblockA{\\Networking Laboratory\\  University of Applied Sciences of Southern Switzerland (SUPSI)\\
%%Manno, Switzerland\\
%%firstName.lastName@supsi.ch}
%%\and
%\IEEEauthorblockN{Alan Ferrari, and Silvia Giordano}
%\IEEEauthorblockA{\\Institute for Information System and Networking\\University of Applied Sciences of Southern Switzerland (SUPSI)\\
%Manno, Switzerland\\
%alan.ferrari@supsi.ch}
%}

\maketitle

\begin{abstract}
Nowadays, privacy has become a very serious issue with smart and mobile platforms. Users tend to allow intrusive apps access much sensible information without really knowing the potential threats. To solve this issue several solutions (e.g. GDPR) have been provided. Our claim is that the users currently are not sufficiently involved in this process for being able to use such solutions. To do this we developed an application that provides a form of awareness to the users and we asked them to reply a set of questions. Our conclusions are that users must be better informed of the risks and value of their personal information.
 \end{abstract}

\section{Introduction}
In the past few years, smartphone applications have
contributed greatly to the quality of users experience in their smart devices 
 allowing them to reach a huge set of functionalities
even bigger when the user?s personal information are used  At the
same time, however, many applications fail to guarantee any
kind of user privacy; this is especially true for Android\footnote{http://www.android.com}
smartphones where the Application Market is self-regulated.

The mechanism that Android offers to increase
user privacy is the use of system permissions: every
time an app is downloaded from the market and the installation
process is started, Android controls the app?s permissions and
then asks the user whether she is willing to allow the app to
use that set of permissions. If the user does not agree, the
installation process is interrupted. This system allows users to
know what information is used and which entity processes
it, but it does not indicate when and how the information
is used inside an application. For instance, if an application
proposes a game that uses sensor data and secretly records
it, the sensor data could be used for malicious intents 
 without the user?s knowledge.

Novel regulations (i.e. GDPR) requires that the user is informed about the data access and collection by the app. However, it is our claim that users are not informed enough to decide whether a data access in an app is a potential threat or not.

With this work, we perform a measurement of the user's impact of a users awareness campaign about the potential risks of the app he/she has installed in their own smart devices.

\begin{figure*}
\centering
\begin{subfigure}{.5\textwidth}
  \centering
  \includegraphics[width=.8\linewidth]{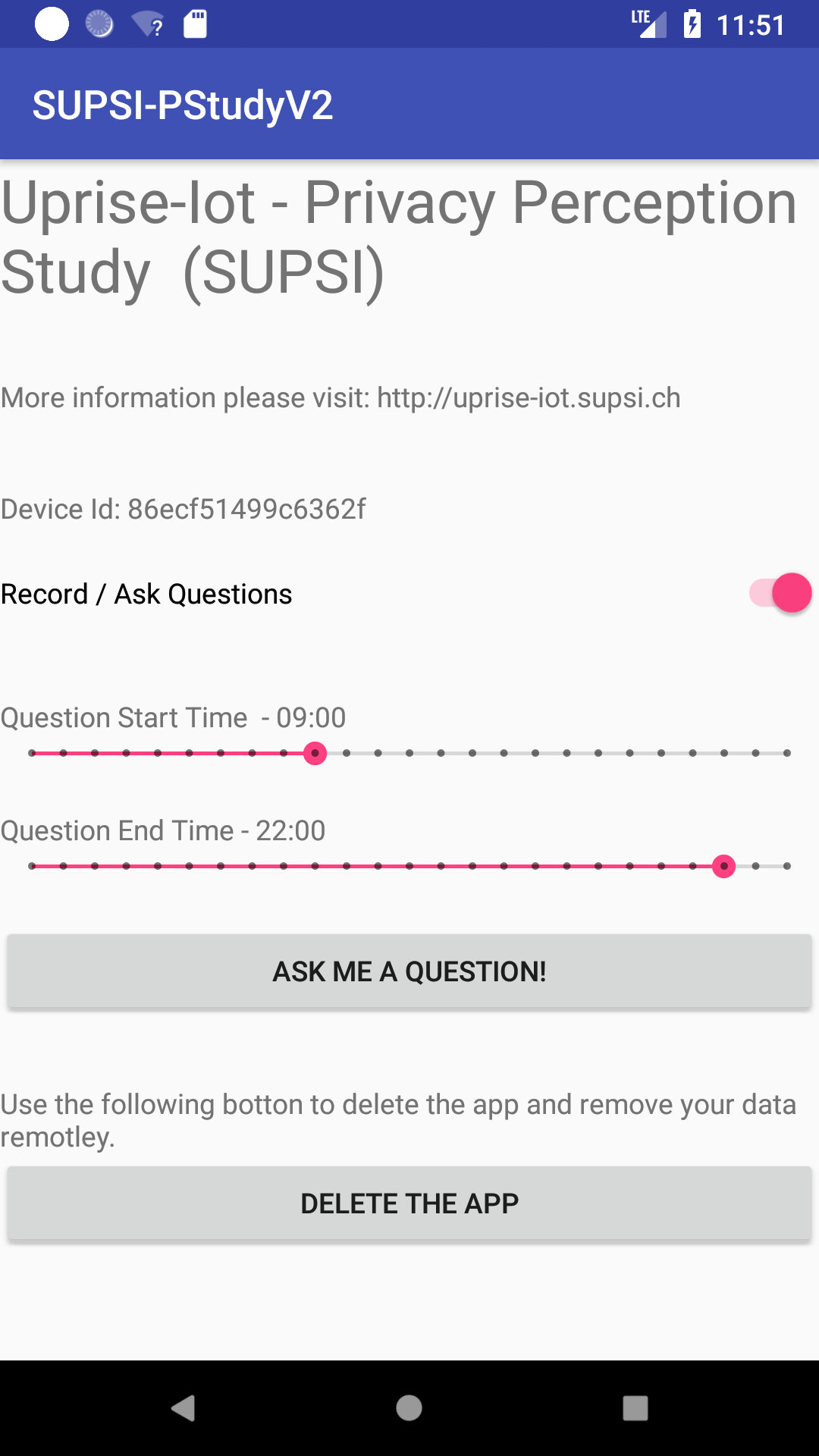}
  \caption{App Main Windows, it allows the user to configure the app based on her personal needs. A delete button allows the user to delete the app and also erase all her data remotely in order to maintain full anonymization. }
  \label{fig:sub1}
\end{subfigure}%
\begin{subfigure}{.5\textwidth}
  \centering
  \includegraphics[width=.8\linewidth]{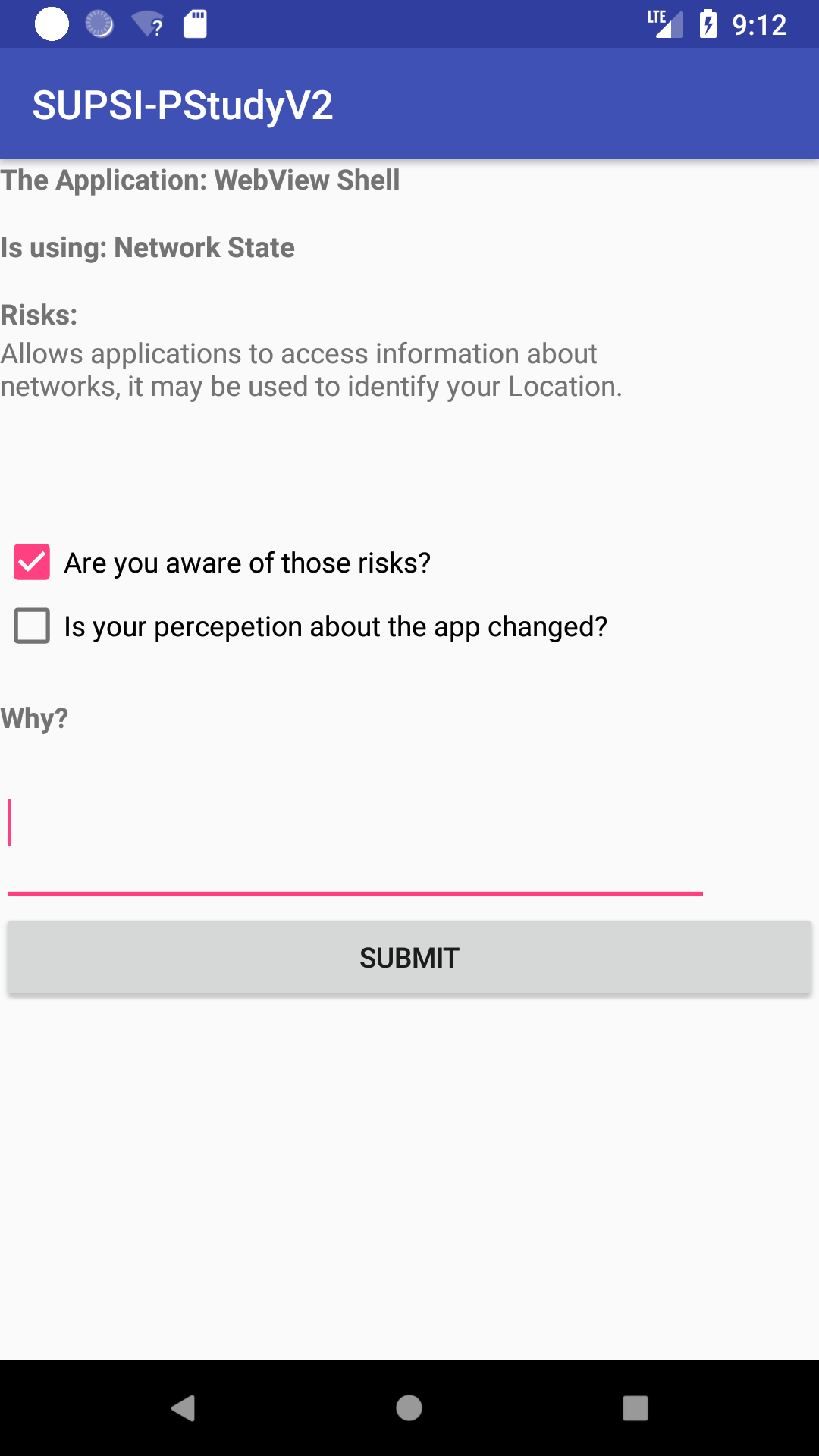}
  \caption{Question Windows, it shows the user a set of information about the behavior of the app and it asks a set of questions to the user. }
  \label{fig:sub2}
\end{subfigure}
\caption{PrintScreen of the two key windows of the application.}
\label{fig:application}
\end{figure*}

We developed an Android application that monitors the behavior of the other apps installed on the user's device. The app also provides two distinct interaction with the device owner:
\begin{itemize}
\item  It informs the owner about a potential risk a given app has considering a specific permission.
\item It asks the user if he/she knows those risks and (in case he didn't) if his/her perception of the app is changed. 
\end{itemize}

We perform the first round of experiments with a selected group of 17 users in order to obtain a first idea of the data we will be able to collect on a larger scale.

\section{Related work}
Privacy has been a hot topic in several domains and has been approached in different ways.

In the context of smart and mobile devices, several key works have been proposed during the past years.  Langheinrich et. al. in \cite{langheinrich2001privacy} defines a set of challenges that any  Ubiquitous/Pervasive application must satisfy in order to be privacy-save. They can be resumed in the following list: 
\begin{itemize}
\item Principle of Openness or simply Notice: users should be aware of the nature of the data shared.
\item  Choice and Consent: where users can choose to offer that information to the requester and they have to give explicit consent.
\item Anonymity and Pseudonymity: Anonymization can be defined as the state of being not identifiable within a set of subjects.
\item Proximity and Locality: in essence it expresses the fact that information must not be disseminated indefinitely, not even across a larger geographic boundary.
\item Adequate Security: Network and disk security are fundamental.
\item Access and Recourse: provide mechanisms for access and regulation. Eventually, also penalties if someone break the rules.
\end{itemize} 
%%%
%Marc Langheinrich. Privacy by design?principles of privacy-aware ubiquitous systems. In International conference on Ubiquitous Computing, pages 273?291. Springer, 200
%
Notable work on smart devices has been performed by  William et. al. on \cite{enck2014taintdroid} provides TaintDroid,
a tool to track and monitor sensible information inside an
application. The idea is to extend and mark each point where
sensible information is used inside the application and track
such points at runtime. It is shown that there exist 68 instances
of information misuses in 20 out of 30 popular applications
downloaded from the market, thus confirming the need for
efficient and easily deployable privacy-preserving techniques.

A possible solution to this problem has been proposed by Ferrari et. al. in \cite{ferrari2015managing} where authors propose Mockingdroid. 
Mocking is a traditional technique in software testing;
its main goal is to mimic the real object behavior in a
controllable way. Recently, mocking techniques have been used
in mobile environments to increase the user privacy and their goal is to allow users to select the kind of information they want
to pass to the application (if real or randomly generated).
The Mockingbird framework is a solution to mocking
that uses recorded context-traces instead of randomly generated
data, which is easily detected by applications. We also propose
a flexible methodology to mock an Android application that
does not require any changes at the operating system level and
at the virtual machine level. Mockingbird is a very promising
solution; we are currently testing its performance and increasing
its functionality

Many of those principles are nowadays included in the law, for instance, the European Data Protection Law  (GDPR) \footnote{https://www.eugdpr.org/} forces the concepts of ?Privacy by Design? and ?Privacy by Default? in its regulation.  Privacy By Design means data protection through technology design" in other words the fact that data protection techniques are already integrated into the data processing with final goal minimize privacy risks through technical and governance controls. 
Privacy by default means that when a system or service includes choices for the individual on how much personal data he/she shares with others, the default settings should be the most privacy-friendly ones.

Even though a valid set of legal tools are nowadays available to the user is our claim that the lack of knowledge of the potential risks related to data access in mobile smart devices makes them less effective.  To this extent, we provide the following study that has a final goal to demonstrate that users must be better informed. 

\section{Study Definition and data collection}
Our goal is to study the privacy perception in the device owned by a group of selected users. To this extent, we develop and Android Application \ref{fig:application} that has as goal the measurement of the key elements: 
\begin{itemize}
\item \textbf{The "privacy level" of the apps installed on the users' device: } we collect the Permission granted to the app by the user and classify them according to the level of risk.
\item \textbf{The level of user awareness:} periodically we show to the user a potential risk connected to the data access for a given app (if there is any) and ask if is aware of that and if with this information his/her perception of the app is changed (in terms of safety).
\end{itemize}

\subsection{Security-Percepetion App}
The app is built for the Android operating system and it is composed of three key entities:
\begin{itemize}
\item \textbf{Background Service: } it collects the following information:
\begin{itemize}
\item  List of the application installed in the devices and the permissions allowed by the users.
\item Application in execution in a specific time interval.
\item System status (CPU / Memory) and context information (location and user's activity)
and it triggers the request for users input.
\end{itemize}
\item \textbf{Main Window \ref{fig:sub1}: } it allows the user to configure the system settings. In other words, the users is able to decide whether to collect information or not and the time he/she are willing to answer our questions.
This activity also allows the user to delete the application and all the information we store from him in a privacy-safe manner.
\item \textbf{Awareness Window \ref{fig:sub2}: } periodically the user receives an information about the behavior of one of the applications installed on his/her phone and, afterward,  he/she also receives the request to answer a question about its perception on a given application installed on his/her phone.  
\end{itemize}
All the information are anonymized and the user is only identifiable by its unique device id. The delete procedure is also secure because  we allowed the user to delete all his/her information by the app thus without making any connection between user and user-id.

\begin{figure}
\centering
\includegraphics[scale=.6]{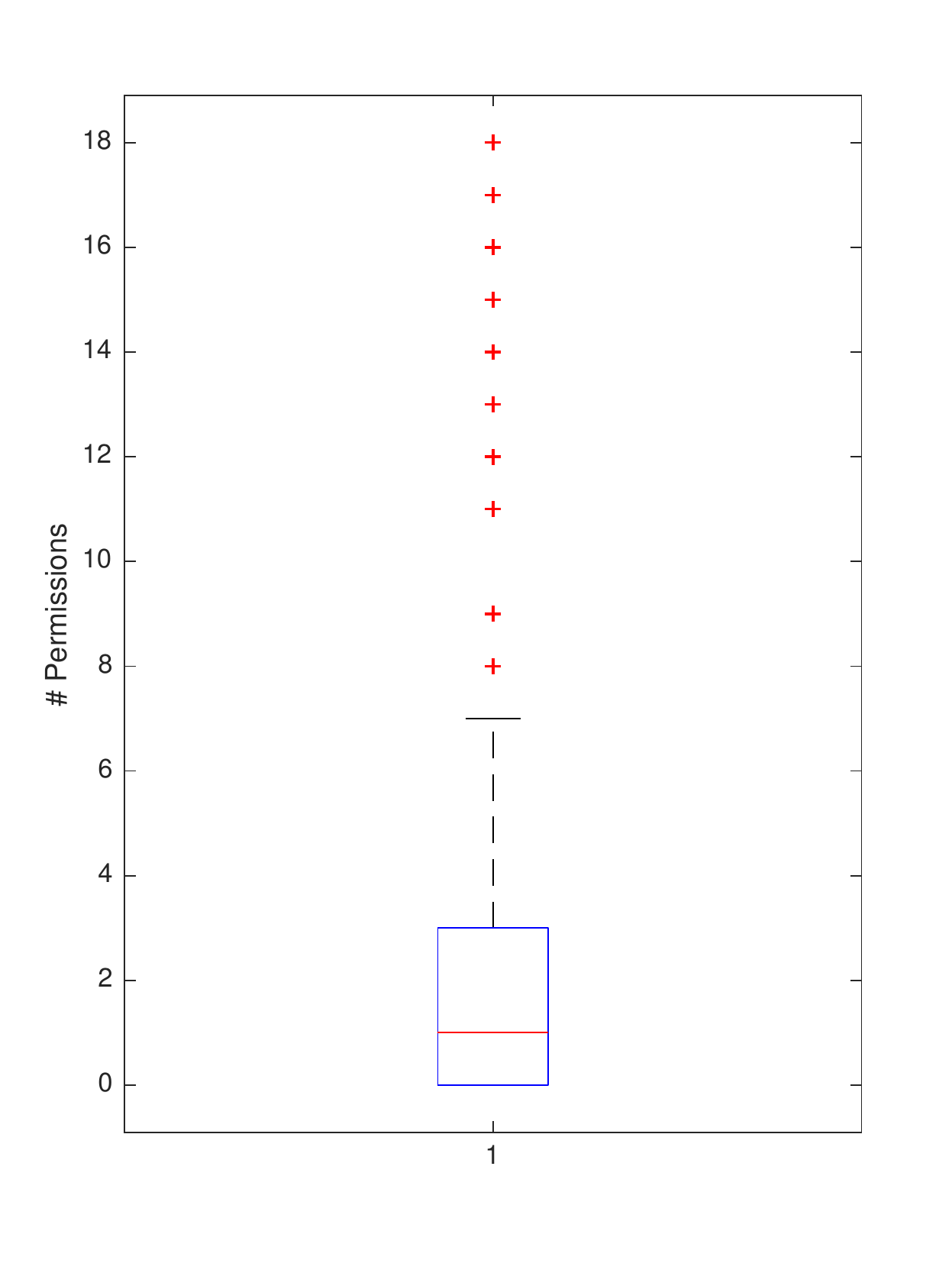}
\caption{Distribution of the number of permissions per app classified as dangerous.}
%%%%
\label{fig:perm}
\end{figure}

During the first run, we ask the user their profile in the form of the age range, gender, and IT knowledge.

\paragraph{To make the user aware} we focus our attention on the permissions the user grant to the application on its phone. To define the risk of the permission we use the categorization provided by  the Android  developers team
\footnote{https://developer.android.com/reference/android/Manifest.permission.html}  where a pre-defined set of permissions are labeled as \textbf{'Protection level: dangerous'}. In order to give an awareness to the user, we show them the application, the permissions it access and a text that describes the overall goal of that permission and the potential risks. 

\begin{figure}
\centering
\begin{subfigure}{.5\textwidth}
  \centering
  \includegraphics[width=1\linewidth]{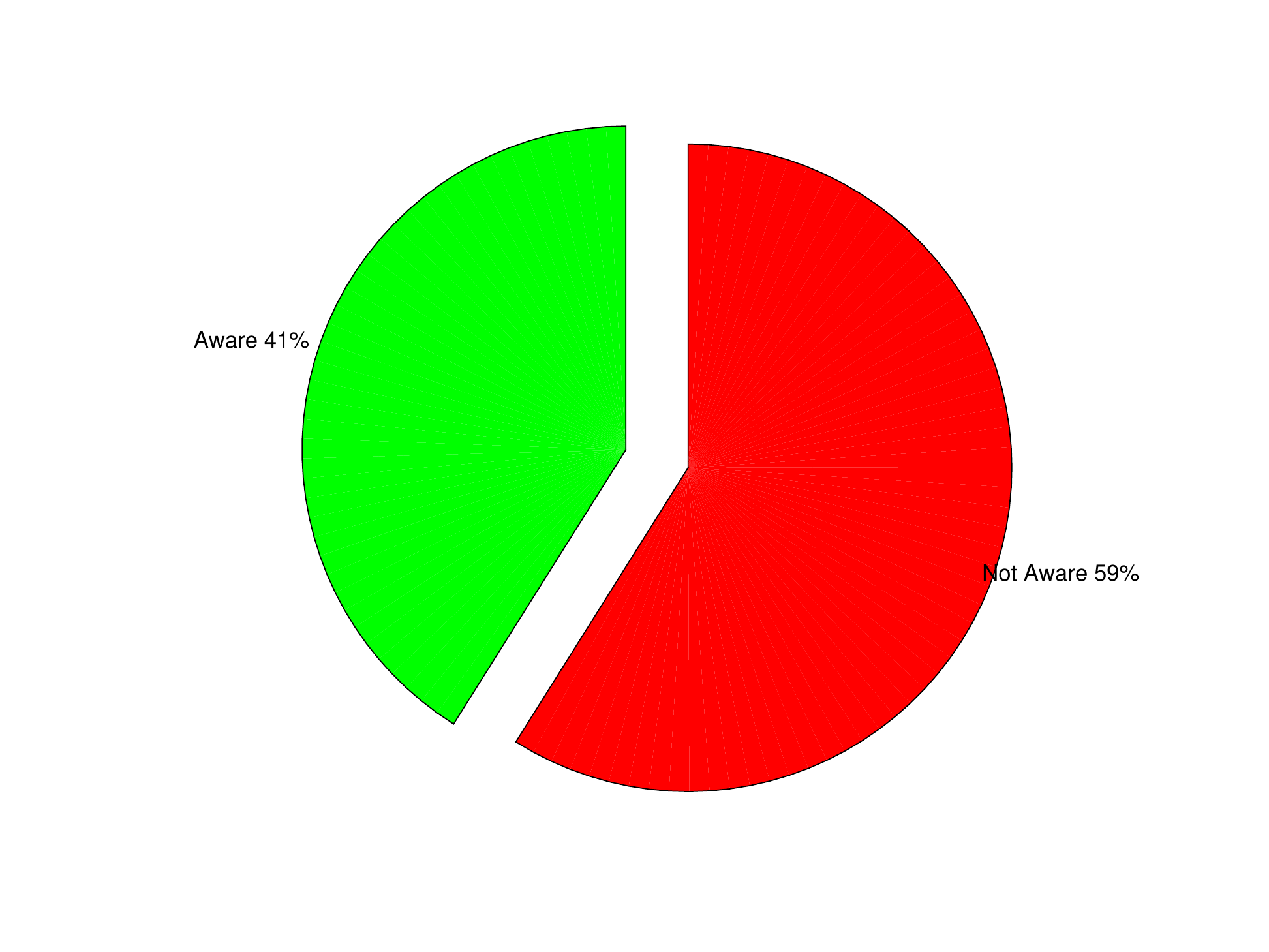}
  \caption{Answers on which the user is aware of the app access [\%]}
  \label{fig:aw}
\end{subfigure}
\begin{subfigure}{.5\textwidth}
  \centering
  \includegraphics[width=1\linewidth]{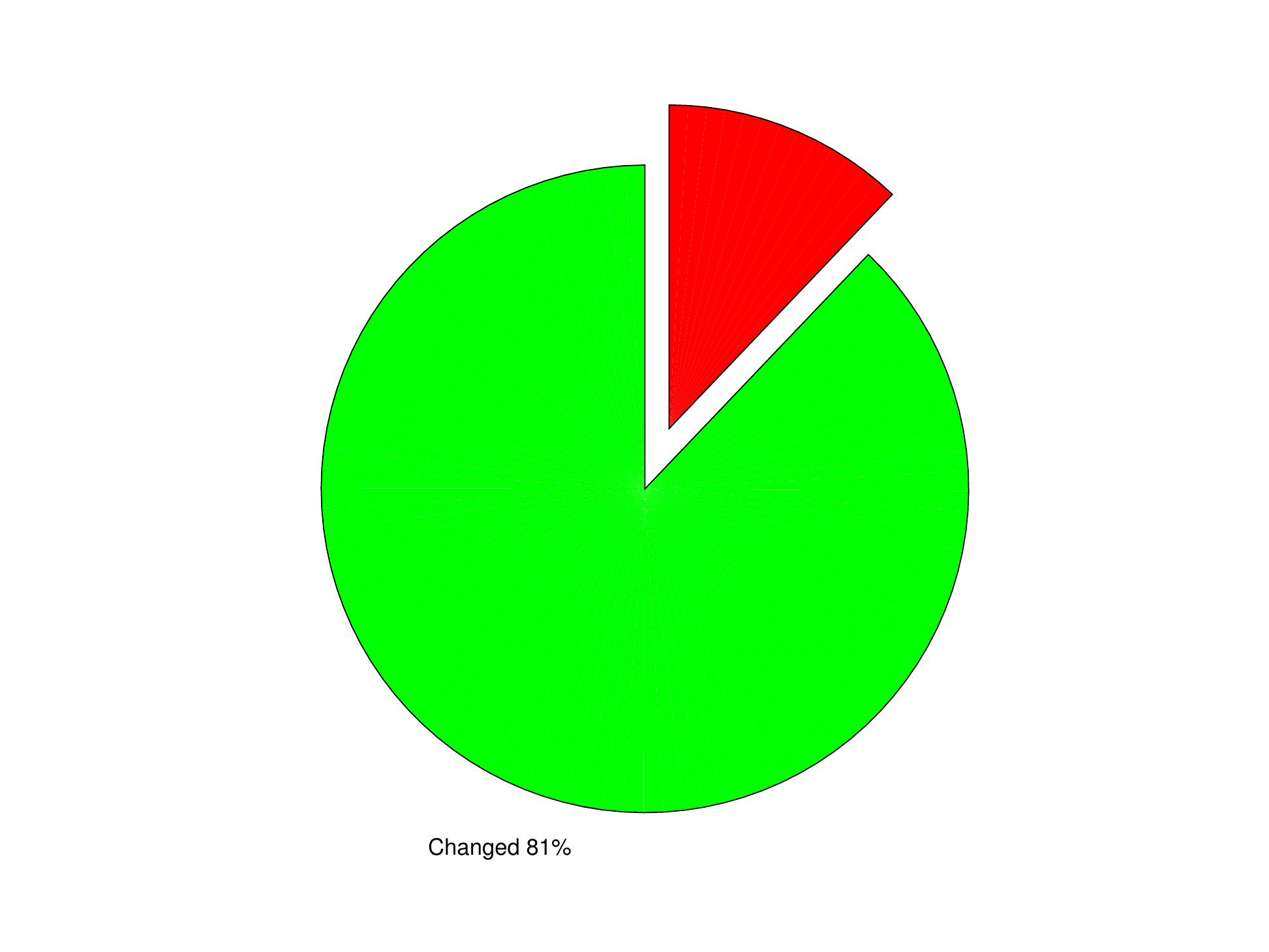}
  \caption{Changes in perception after the awareness phase [\%]}
  \label{fig:achanges}
\end{subfigure}%
\caption{Aggregate outcome of the users' answers.}
\label{fig:outcomes_user}
\end{figure}

\paragraph{To measure the impact of our awareness} we ask to every time he sees the awareness two key questions:
\begin{itemize}
\item Do you know that this app access that information?
\item In case you don't know; has your perception of the app changed?
\end{itemize} 
These questions allow us to measure if the user is aware of the app dangerous data access (maybe because the app need that information) and also to measure the impact of our awareness with the help of the second question.

\section{Experiments and key findings}
Our claim is that the users are not enough aware of the behavior of their application. Users pay very little attention to the data request made by the app, many reasons are behind this behavior but we believe that the lack of knowledge of the risks is one of the key elements. 

To prove our claim we ask 17 subjects to use our app and answer as much question as they can. 
The user profiles are shown in Table \ref{table:1}, we see that they span across different ages, gender and it knowledge making the set we collected heterogeneous. 
		
The next analysis we performed is connected to the dangerous permissions required by the app installed on the users' phones. In Figure \ref{ifg:perm} is shown the distribution. We clearly see that the majority of the app requires numbers of permission between 0 and 7 with mean 4.  However, there are a group of outliers that requires an astonishing number of dangerous permission up to 16.

The key outcome of this first analysis is that the privacy of user sensible information may be endangered by the majority of application he/she has installed in his/her mobile device.

As said before, to determine if the user is aware of the potential risks the app he/she has installed in his/her phone we provide an awareness in the form of a notification of a potential risk in a given application when we notice the application is under execution. The notification contains the app name, the permission it access and a list of potential risks connected to this access.

\begin{table}[h!]
\centering
\begin{tabular}{l l l l} 
\textbf{Age} &  31 - 40: 8 & 41 - 50: 4 &   18 - 30: 7\\
\textbf{Gender} &  Female: 6 &  Male: 11 & \\
\textbf{It } & Advanced: 11 &  Some: 5 & None: 3 \\
\end{tabular}
\caption{User profile categorizations.}
\label{table:1}
\end{table}
To measure the impact of the awareness we asked the user to reply to key questions; if he/she was aware of the app dangerous accesses, and, in case he was not aware of that, if his/her perception of the app has changed. We clearly see that in the majority of the cases (more than 60\%) the user \textbf{has no idea} of the risks connected to that app.

Another important measure is to determine if the awareness has an impact on the users; to do so we asked the users if his/her perception has changed in the case of a successful awareness. The outcome is presented in Figure \ref{ig:achanges}; there, the results show that in the circumstance the user is well informed he/she is able to recognize the potential threats in the app and therefore he/she changed his/her perception of the app.

\section{Conclusions and future work}
With this work, we provide a first study on the perception of privacy in users of mobile smart devices.  We develop an app that studies the permissions of the other apps installed on a smart device and if those app access data that may lead to privacy damages we informed the users with our awareness campaign. To measure the impact of the awareness we asked two questions to the users that let us understand if the users are aware or not to the risks connected to the apps data access and, in case he/she is not aware of that we asked if his/her perception of the app is changed. 

Currently, we made an experiment with 17 users to have a first idea of the data we may be able to collect. The user has been selected to be heterogeneously and from the results, we clearly depict the fact that a better awareness must be presented to the users before providing technological and legal solutions. 

As future work, we plan to extend our analysis including more users and also study the relationship between user profile and impact of the awareness by collecting the information about changes in permissions grants and install/uninstall of the application in the users' devices.

\bibliographystyle{unsrt}	% (uses file "plain.bst")
\bibliography{af-percepetionV1}

\end{document}